\documentstyle[preprint,revtex,eqsecnum]{aps}
\begin{document}
\draft
\preprint{UAHEP-936}
\begin{title}
Complete Semiclassical Treatment \\
of the \\
Quantum Black Hole Problem
\end{title}
\author{B. Harms and Y. Leblanc}
\begin{instit}
Department of Physics and Astronomy, The University of
Alabama,\\
Tuscaloosa, AL 35487-0324
\end{instit}
\begin{abstract}
Two types of semiclassical calculations have been used to
study quantum effects in black hole backgrounds, the WKB and
the mean field approaches.  In this work we systematically
reconstruct the logical implications of both methods on
quantum black hole physics and provide the link between
these two approaches.  Our conclusions completely support
our previous findings based solely on the WKB method:
quantum
black holes are effectively p-brane excitations and,
consequently, no information loss paradox exists in this
problem.
\end{abstract}

\section{Introduction}

In four or higher dimensions two types of semiclassical
calculations, both applied during the 1970's, have been used
 vin the study of quantum effects in black hole backgrounds.
The first one is the familiar Euclidean path integral
formulation of general relativity (GR) in the semiclassical
WKB approximation.  The standard interpretation of this
quantity by Hawking and others \cite{hawk1,hawk2,gibb,bek}
was taken to be the canonical
partition function of a gas of black holes of equal mass in
thermal equilibrium at inverse Hawking temperature
$\beta_H$, the latter quantity being related to the mass of
the black hole.  A second type of calculation aimed at
studying the quantum field theory of particles immersed in
the classical black hole background\cite{bir}.  It was found
that the
effect of the non-trivial topology of the black hole
spacetime,
due to its horizon, was to modify the quantization procedure
by doubling the number of degrees of freedom of the given
quantum fields as compared to the expected number in a
toplogically trivial, horizonless, spacetime.  This quantum
field theory then acquired a mathematical structure
analogous to flat space quantum field theories at finite
temperature (e.g. the thermofield dynamics
formalism\cite{umez}).
Furthermore, the temperature of the gas of particles in the
black hole background agreed with that found in the WKB
calculation.  It seemed as though thermodynamics was
naturally coming out of quantum black hole theory.  This led
Hawking to propose a new set of physical laws -- black hole
thermodynamics.

Over time however, an increasing number of theorists raised
concerns over the potential violations of quantum mechanics
implied by this new black hole thermodynamics.  In
particular, unitarity of time evolution would be broken as
particles in pure states originally absorbed by the black
hole would tunnel out of it and end up in the above
described thermal state.  Inconsistency of the thermal
picture for black holes could also be detected in the form
of a negative canonical specific heat, a necessarily
positive definite quantity.

In this work no toy model (whether in 2 or other dimensions)
is used.  We address the real problem.  The purpose of this
paper is to systematically re-assemble our knowledge of
quantum black hole physics in 4 or higher dimensions solely
on the basis of the two calculational methods described
above for dealing with quantum effects in gravity problems.
Conjectures are eliminated as much as possible.  A clear
picture of the nature of quantum black holes finally
emerges, a picture in agreement with the currently known
laws of physics.  Such a picture is the one described by the
present authors in previous
publications\cite{hl1,hl2,hl3,hl4}.  What is new in
this paper is the consistent incorporation of the second
computational semiclassical method and the elucidation of
its role in the understanding of the overall picture.  This
was not included in our previous work.

\section{The WKB Method}

In the 1970's one of the very few methods available to deal
with the problem of extracting useful and {\it finite}
information from quantum gravity theory was the WKB
semiclassical approximation\cite{hawk1,hawk2,gibb,bek}.  It
was
pioneered by Hawking,
among others, in the context of black holes.  This technique
provides an approximate evaluation of the path integral of
Euclidean field theories by finding saddle points.  The
result is generically given by the exponential of minus the
Euclidean action (over $\hbar$) evaluated at classical
solutions (instantons) of the Euclidean field equations,
multiplied by an overall factor which is determined by the
quantum corrections.  It is a clearly nonperturbative
calculation.

Because of its connection with instantons, the WKB
approximation to the path integral provides a useful formula
for the tunneling probability per unit volume of quantum
particles in barrier penetration problems.

In the case of a $D$-dimensional Schwarzschild black hole an
analytical continuation from real to imaginary time must be
performed to evaluate the above tunneling probability across
the horizon.  In this process a conical singularity in the
Euclidean spacetime will develop, however.  The situation is
remedied by further demanding that the imaginary time
dimension be constrained to form a compact circle with
circumference $\beta_H$, the well-known Hawking inverse
``temperature''.  The gravitational instantons that are the
Euclidean black holes are therefore in fact periodic
instantons\cite{gross}.

Since the tunneling probability is an effective measure of
the ratio of a single particle state having escaped the
black hole to the number of quantum states inside the black
hole, we therefore arrive at the following WKB formula for
the black hole {\it quantum} degeneracy of states $\rho(m)$
at mass level
$m$,
\begin{eqnarray}
\rho(m) {\sim} c\; e^{S_E(m)/\hbar} \; ,
\end{eqnarray}
where $c$ represents the quantum field theoretical
corrections and $S_E$ is the Euclidean action of the
Euclidean black hole (evaluated from the horizon outward).
The integral over Euclidean time in Eq.(2.1) is to be
performed for a single period.  The picture thus presented
is {\it completely quantum mechanical}.

On the other hand, ever since the mid-fifties when Matsubara
\cite{matsu} proposed his imaginary time formalism for
equilibrium
quantum field theory at finite temperature, the Euclidean
path integral formulation has also been interpreted as the
canonical partition function $Z(\beta)$ of a gas in thermal
equilibrium.  The inverse temperature $\beta_H$ is again the
period of the Euclidean time.  Hawking and
others\cite{hawk1,hawk2,gibb,bek}, in the
mid-seventies, chose this second interpretation of the path
integral in the black hole context, a pioneering effort to
understand the statistical mechanics of these unusual
objects.  From there the {\it statistical} density of states
$\Omega(E)$ of a gas of black holes with average energy $E =
M$ can be found.  The canonical partition function
is written as
\begin{eqnarray}
Z(\beta_H) \sim e^{-S_E(\beta_H)/\hbar} = e^{-\beta_H
F(\beta_H)} \; ,
\end{eqnarray}
where $F(\beta_H)$ is the Helmholtz free energy. The
corresponding entropy is
\begin{eqnarray}
S_H = \beta_H M - \beta_H F(\beta_H) = \beta_H M - S_E \; ,
\end{eqnarray}
and so the statistical mechanical density of states is now
given as,
\begin{eqnarray}
\Omega_H(M) = e^{S_H(M)} \; .
\end{eqnarray}

Let us consider explicitly the simple problem of the
$D$-dimensional Schwarzschild black hole.  The Euclidean
metric
is
\begin{eqnarray}
ds^2 = e^{2\Phi} d\tau^2 + e^{-2\Phi} dr^2 + r^2 d\Omega_{D-
2}^2 \; ,
\end{eqnarray}
where
\begin{eqnarray}
e^{2\Phi} = 1 - \Bigl({r_+\over{r}}\Bigr)^{D-3} \; ,
\end{eqnarray}
and $r_+$ is the horizon radius.

The condition of the vanishing of the conical singularity of
the spacetime (2.5) yields the Hawking inverse temperature
\begin{eqnarray}
\beta_H = {2\pi\over{[e^{\Phi}\partial_re^{\Phi}]_{r=r_+}}}
={4\pi r_+\over{D-3}} \; .
\end{eqnarray}

The Euclidean action for such a $D$-dimensional black hole
has been repeatedly derived.  The result is $(\hbar = 1)$,
\begin{eqnarray}
S_E = {A_{D-2}\over{16\pi}} \beta_H r_+^{D-3} \; ,
\end{eqnarray}
where $A_D$ is the area of a unit $D$-sphere.  The relation
between the horizon $r_+$ and the black hole mass $M$ is
given by
\begin{eqnarray}
M = {(D-2)\over{16\pi}} A_{D-2} r_+^{D-3} \; .
\end{eqnarray}
Therefore,
\begin{eqnarray}
S_E(M) = {\beta_H M\over{D-2}} = \sigma(D) M^{{D-2\over{D-
3}}} \; ,
\end{eqnarray}
where,
\begin{eqnarray}
\sigma(D) = {4^{D-1\over{D-3}}\pi^{D-2\over{D-3}} \over{(D-
3)(D-2)^{D-2\over{D-3}} A_{D-2}^{1\over{D-3}}}} \; .
\end{eqnarray}
The Hawking entropy is now given as
\begin{eqnarray}
S_H(M) = (D-3)S_E(M) = (D-3)\sigma(D) M^{D-2\over{D-3}} \; .
\end{eqnarray}
Therefore,
\begin{eqnarray}
\rho(M) \sim e^{\sigma(D) M^{D-2\over{D-3}}} \; ,
\end{eqnarray}
and
\begin{eqnarray}
\Omega_H(M) \sim \rho^{D-3}(M) \; .
\end{eqnarray}
According to the last equation, although both the quantum
and Hawking density of states agree in 4 dimensions the
thermodynamical interpretation predicts a vastly enhanced
number of states in higher dimensions as compared to the
quantum interpretation.

Let us now analyze very closely the implications of the
thermodynamical interpretation.  First as is clear from
Eq.(2.2), the canonical partition function is finite in the
WKB approximation.  On the other hand, the exact partition
function is the Laplace transform of the density of states
(2.14).  We get
\begin{eqnarray}
Z(\beta) = \int_0^{\infty}dE e^{-\beta E} e^{(D-
3)\sigma(D)E^{D-2\over{D-3}}} \; .
\end{eqnarray}
For $D\geq4$, the above expression diverges badly for all
$\beta$, implying the non-existence of the canonical
partition function for all temperatures.  The non-existence
of the canonical ensemble can also be seen from the
calculation on the canonical specific heat.  It is found to
be negative.  The saddle point approximation therefore must
fail and the canonical and microcanonical ensembles are not
equivalent.  The study of the statistical mechanics of black
holes must
proceed in the microcanonical ensemble.  The result of
Eqs.(2.2) and (2.14) must therefore be wrong.  Black hole
thermodynamics is simply not a viable option.  Eq.(2.15)
gives the proof.

This leaves us with only one alternative, the fully quantum
interpretation of the WKB formula as the tunneling
probability, yielding directly the {\it quantum} black hole
degeneracy of states Eqs.(2.1) and (2.13).

Comparison of the degeneracy of states (2.13) with those of
known non-local quantum theories yields immediately the
classification of $D$-dimensional Schwarzschild black holes
as the quantum excitation modes of a $\Bigl({D-2\over{D-
4}}\Bigr)$-brane.  Black holes are therefore fully
elementary particles.  Massless particles such as photons
may then be regarded as extreme quantum black holes with a
horizon of zero radius.

The study of the statistical mechanics of a gas of such
objects in the microcanonical ensemble (the unique approach)
reveals their conformal nature through two characteristics,
the validity of the statistical bootstrap property and the
duality (crossing symmetry) of the S-matrix.  The
equilibrium state of a gas of $N$ black holes is found
\cite{hl1,hl2,hl3,hl4} to be
the one for which there is a single very massive black hole
in the gas and $(N-1)$ massless others, a state very far
from thermal equilibrium.  The microcanonical specific heat,
of course, is negative.  The above fully quantum picture
therefore resolves completely the so-called information loss
paradox.

\section{Mean-Field Theory}

The results of the preceding section were arrived at within
the semiclassical WKB approximation.  Besides the WKB method
there is another semiclassical technique (mean field theory)
which aims at
quantizing field theories in the classical black hole
background\cite{bir}.

The problem of quantization can be approached from the
viewpoint of scattering theory, in which quantum fields are
scattered off the black hole horizon.  Because of the
horizon, two causally disconnected spacelike regions coexist
and field quantization in each sector makes use of a
different Hilbert (Fock) space.  When quantum fields scatter
off the horizon, mixing occurs between the corresponding two
sets of modes.  This is expressed mathematically as a
Bogoliubov transformation.  This doubling of the number of
degrees of freedom of the theory due to the non-trivial
topology of the black hole spacetime, turns out to bear
considerable resemblance to the mathematical structure of
modern field theories at finite temperature (e.g. the
thermofield dynamics formalism\cite{umez}).  In particular,
the vacuum
state for the outgoing particle can be formally written as
follows,
\begin{eqnarray}
|out,0> = Z^{-1/2}(\beta)\sum_{n=0}^{\infty} e^{-\beta
n\omega
/2}|n>\otimes |\tilde{n}> \; ,
\end{eqnarray}
in which $|n>$ and $|\tilde{n}>$ are the Fock spaces of the
two causally disconnected regions (the observer sees only
$|n>$ directly).

It follows that the physically observable correlation
functions are those obtained by making use of the vacuum
(3.1).  It is easy to show that any expectation value with
respect to the above vacuum is equivalent to a statistical
average in the canonical ensemble.  For any observable $A$
we have
\begin{eqnarray}
<out,0|A|out,0> = \sum_{n=0}^{\infty} e^{-\beta n\omega}
<n| A|n>/Z(\beta) \; ,
\end{eqnarray}
where the partition function $Z(\beta)$ is given by
\begin{eqnarray}
Z(\beta) = \sum_{n=0}^{\infty} e^{-n\beta\omega} \; .
\end{eqnarray}

The inverse temperature $\beta$ is
\begin{eqnarray}
\beta = {2\pi\over{\kappa}} \; ,
\end{eqnarray}
where $\kappa$ is the surface gravity of the black hole.
Therefore $\beta = \beta_H$, in agreement with the WKB
result.

A direct consequence of Eq.(3.2) is the fact that the mass
$m$ particle number density has the following Planckian
(thermal) distribution for an observer outside the event
horizon
\begin{eqnarray}
n_k(m;\beta_H) = {1\over{e^{\beta_H\omega_k(m)} - 1}} \; ,
\end{eqnarray}
a fact interpreted as a loss of information during the
scattering process with the black hole.  Since the
scattering process started with particles in a pure state (
the ``in'' vacuum given by $|in,0> = |0>\otimes
|\tilde{0}>$), Eq.(3.5) implies a loss of unitarity during
the scattering process, a violation of quantum mechanics (
in the field theory limit),
\begin{eqnarray}
|out,0> = S^{-1}(\beta)|in,0> \; ; \ \ S^{-1} \neq
S^{\dagger}
\; .
\end{eqnarray}

Clearly the results of the previous section show that a
thermodynamical description of quantum black hole physics is
inappropriate.  Thus the legitimate question: what about the
result of Eq.(3.5)?  In the following section we explain how
one reconciles both semiclassical considerations.

\section{Semiclassical Black Holes}

In this section, we explain how the seemingly contradictory
semiclassical results of the preceding two sections
actually do reconcile.

The inescapable conclusion of section II is the non-local
nature
of the semiclassical black holes. On the other hand, the
result of
Eq.(3.5) for the particle number density  is that of a local
field
theory. This result therefore cannot be accepted at face
value and
cannot determine alone the true vacuum of our black hole
problem.

Clearly, and it is here that back reaction effects start
entering
the picture, one needs to consider the {\it total} particle
number
density of all the particle modes of the full non-local
quantum
gravity theory. For a complete treatment, Eq.(3.5) must then
be
replaced by the following general expression,
\begin{eqnarray}
{n_k}(\beta_H)\;=\;\int_0^{\infty}dm\,{\rho(m)}{n_k}(m;
{\beta_H})\;.
\end{eqnarray}
Eq.(4.1) leads to the following canonical partition
function,
\begin{eqnarray}
Z(\beta_H)\;=\;\exp\Bigl(-{V\over{(2\pi)^{D-1}}}{\int_{-
\infty}^{\infty}}
{d^{D-1}}{\vec k}{\int_0^{\infty}}dm\,{\rho(m)}\ln{[1\,-\,
e^{-{\beta_H}{\omega_k}(m)}]}\Bigr)\;,
\end{eqnarray}
where ${\omega_k}(m)\,=\,\sqrt{{{\vec k}^2}+{m^2}}$.

The ``thermal vacuum'' Eq.(3.1) is now generalized as
follows,
\begin{eqnarray}
|out,0>\;=\;Z^{-1/2}(\beta)\,[{\prod_{m,k}}\ \
{\sum_{{n_{k,m}}=0}^{\infty}}] \prod_{m,k}\,e^{-
{\beta\over 2}{n_{k,m}}{\omega_{k,m}}}
|{n_{k,m}}>\otimes|{{\tilde{n}}_{k,m}}>\;.
\end{eqnarray}
Recalling that
\begin{eqnarray}
Z({\beta_H})\;=\;{\int_o^{\infty}}dE\,e^{-
{\beta_H}E}{\Omega(E)}\;,
\end{eqnarray}
and comparing the above equation with Eq.(4.2), one finally
arrives at Hagedorn's old self-consistency
condition\cite{hage},
\FL
\begin{eqnarray}
\exp\Bigl(-{V\over{(2\pi)^{D-1}}}{\int_{-\infty}^{\infty}}
{d^{D-1}}{\vec k}{\int_0^{\infty}}dm\,{\rho(m)}
\ln{[1\,}&-&{\,e^{-{\beta_H}{\omega_k}(m)}]}\Bigr)\nonumber
\\
&=&{\int_0^{\infty}}dE\,e^{-{\beta_H}E}{\Omega(E)}\; .
\end{eqnarray}
We are seeking solutions obeying the statistical bootstrap
requirement,
\begin{eqnarray}
{{\rho(E)}\over{\Omega(E)}}\;\rightarrow\;1\;\;;\ \
(E\rightarrow\infty)\;.
\end{eqnarray}
As is well known, string theories are the only possible
solutions
of the above self-consistency condition,
\begin{eqnarray}
\rho(m)\;\sim\;e^{bm}\;\;;\ \ (m\rightarrow\infty)\;,
\end{eqnarray}
provided ${\beta_H}>b$, where $b^{-1}$ is the so-called
Hagedorn temperature.

The WKB results of section II exclude however string
theories as
quantum black hole theories. Black holes are p-brane
excitations
with $p={{D-2}\over{D-4}}$ and so $p>1$. Black hole
solutions
are therefore excluded as solutions of the conditions (4.5)
and (4.6)
as
they yield an infinite canonical partition function.
Therefore,
the thermal vacuum Eq.(4.3) is the {\it false vacuum}. Again
one
finds that thermal
equilibrium is alien to quantum black hole physics.
Semiclassical
quantization as discussed in the previous section is the
wrong
starting point for field quantization in a black hole
spacetime.
Note that to arrive at this conclusion, one needs consider
the
full non-locality of the quantum gravity theory by including
the effects of all the excitation modes (back reactions) of
the theory.

The obvious next question is how to quantize fields in black
hole
backgrounds. This is not an easy question to answer.
However, it may
be possible that, recalling the nature of the non-thermal
equilibrium
state of a gas of black holes, one might need some kind of
generalization of usual (``canonical'') quantum field
theory to the so-called microcanonical quantum field theory.

\narrowtext
\section{Conclusion}

In this work, we presented a complete treatment of the
semiclassical
approaches to quantum black hole physics.

In section II, we reviewed the resolution of the so-called
information
loss paradox, as provided in our earlier
works\cite{hl1,hl2,hl3,hl4,hl5}. In section IV, we
provided the solution to the thermal spectrum problem by
taking full account of the non-locality of the quantum
theory of gravity. Only
then could the results of both semiclassical calculations be
brought to agreement.

Again all our considerations are consistent with the view
that quantum
$D$-dimensional black holes are excitation modes of ${{D-
2}\over{D-4}}$-
branes.

Of course a deeper understanding of these results remains
the subject
of future endeavors.

\acknowledgments

This research was supported in part by the U.S. Department
of Energy under Grant No. DE-FG05-84ER40141.

\end{document}